\documentclass[journal]{IEEEtran}

\ifCLASSINFOpdf
\else
\fi

\usepackage{booktabs}
\usepackage{graphicx}
\usepackage{tipa}
\usepackage{tabularx}
\usepackage{float}
\usepackage{subfig}
\usepackage{url}

\usepackage{breakurl}
\usepackage[breaklinks]{hyperref}

\begin{document}


%
%

\title{Robust Estimation of Hypernasality in Dysarthria with Acoustic Model Likelihood Features}

\author{Michael~Saxon$^1$, 
 Ayush~Tripathi$^2$, 
 Yishan Jiao$^3$, Julie Liss$^3$,
        and~Visar~Berisha$^{1,3}$\\\vspace{1mm}
    $^1$School of Electrical, Computer, and Energy Engineering, Arizona State University\\
    $^2$Department of Electrical and Electronics Engineering, Visvesvaraya National Institute of Technology\\
    $^3$College of Health Solutions, Arizona State University
    
        \thanks{This work is partially supported by National Institutes Of Health Grant 5R01DC006859-14.}

}

\IEEEaftertitletext{\vspace{-2.2\baselineskip}}

%
%

\markboth{}%
{Saxon \MakeLowercase{\textit{et al.}}: Robust Estimation of Hypernasality in Dysarthria with Acoustic Model Likelihood Features}

\maketitle

\begin{abstract}
Hypernasality is a common characteristic symptom across many motor-speech disorders. For voiced sounds, hypernasality introduces an additional resonance in the lower frequencies and, for unvoiced sounds, there is reduced articulatory precision due to air escaping through the nasal cavity. However, the acoustic manifestation of these symptoms is highly variable, making hypernasality estimation very challenging, both for human specialists and automated systems. Previous work in this area relies on either engineered features based on statistical signal processing or machine learning models trained on clinical ratings. Engineered features often fail to capture the complex acoustic patterns associated with hypernasality, whereas metrics based on machine learning are prone to overfitting to the small disease-specific speech datasets on which they are trained.  Here we propose a new set of acoustic features that capture these complementary dimensions. The features are based on two acoustic models trained on a large corpus of healthy speech. The first acoustic model aims to measure nasal resonance from voiced sounds, whereas the second acoustic model aims to measure articulatory imprecision from unvoiced sounds. To demonstrate that the features derived from these acoustic models are specific to hypernasal speech, we evaluate them across different dysarthria corpora. Our results show that the features generalize even when training on hypernasal speech from one disease and evaluating on hypernasal speech from another disease (e.g., training on Parkinson's disease, evaluation on Huntington's disease), and when training on neurologically disordered speech but evaluating on cleft palate speech. 
\end{abstract}


%
\IEEEpeerreviewmaketitle

\section{\label{sec:1} Introduction}

\IEEEPARstart{H}{ypernasality} refers to the perception of excessive nasal resonance in speech, caused by velopharyngeal dysfunction (VPD), an inability to achieve proper closure of the velum, the soft palate regulating airflow between the oral and nasal cavities. Hypernasality is a common symptom in motor-speech disorders such as Parkinson's Disease (PD), Huntington's Disease (HD), amyotrophic lateral sclerosis (ALS), and cerebellar ataxia, as velar movement requires precise motor control \cite{pdhdhn}, \cite{hpark}, \cite{hnat}, \cite{msdb}. It is also the defining perceptual trait of cleft palate speech \cite{kuehn2000speech}. Reliable detection of hypernasality is useful in both rehabilitative (e.g., tracking the progress of speech therapy) and diagnostic (e.g., early detection of neurological diseases) settings \cite{rivera1974}, \cite{chronichead}. Because of the promise hypernasality tracking shows for assessing neurological disease, there is interest in developing strategies that are robust to the limitations of existing techniques---in this work we present automated metrics for hypernasality scoring that are robust to disease- and speaker-specific confounders.

Clinician perceptual assessment is the gold-standard technique for assessing hypernasality \cite{EXTENCE20171014}. However, this method has been shown to be susceptible to a wide variety of error sources, including stimulus type, phonetic context, vocal quality, articulation patterns, and previous listener experience \cite{kent-limits}. Additionally, these perceptual metrics have been shown to erroneously overestimate severity on high vowels when compared with low vowels \cite{kuehn-phon}, and vary based on broader phonetic context \cite{lintz}. Although these difficulties may be mitigated by averaging multiple clinician ratings, this further drives up costs associated with hypernasality assessment and makes its use as a trackable metric over time less feasible.

Automated hypernasality assessment systems have been proposed as an objective alternative to perceptual assessment. Instrumentation-based direct assessment techniques visualize the velopharyngeal closing mechanism using videofluoroscopy \cite{henningssonfluor} or magnetic resonance imaging \cite{kao2008magnetic} and provide information about velopharyngeal port size and shape \cite{bettens}. These methods are invasive and can cause patients pain and discomfort. Alternatively, nasometry measures \textit{nasalance}, the modulation of the velopharyngeal opening area, by estimating the acoustic energy from the nasal cavity relative to the oral cavity with two microphones separated by a plate that isolates the mouth from the nose \cite{pentax}. In some cases, nasalance scores yield a modest correlation with perceptual judgment of hypernasality \cite{brancamp2010relationship, watterson1993relationship}; however, there is considerable evidence that this relationship depends on the person and the reading passages used during assessment \cite{watterson1993relationship}, \cite{sinko2017assessment}. Furthermore, properly administering the evaluation requires significant training and it cannot be used to evaluate hypernasality from existing speech recordings. Thus, the clinician's perception of hypernasality is often the de-facto gold-standard in clinical practice \cite{chapman2016americleft}. 


An appealing alternative to instrumentation-based techniques is the direct estimation of hypernasality from recorded audio. This family of methods aims to measure the atypical acoustic resonance from VPD as an objective proxy for hypernasal speech. Systems that can accurately and consistently rate changes to a patient's hypernasality directly from the speech signal could enable remote tracking of neurological disease progression from a mobile device or home computer \cite{beijer2015asynchronous}.

Previous work in this area can be categorized broadly in two groups: engineered features based on statistical signal processing \cite{orozcorev} and supervised methods based on machine learning \cite{HEGDE2018}. The simple acoustic features fail to capture the complex manifestation of hypernasality in speech, as there is a great deal of person-to-person variability \cite{lohmandercrit}. The more complex machine learning-based metrics are prone to overfitting to the necessarily small disease-specific speech datasets, making it difficult to evaluate how effectively they generalize.

In this paper, we propose an approach that falls between these two extremes. We know that for voiced sounds hypernasal speech results in additional resonances at the lower frequencies \cite{schools}. The acoustic manifestation of the additional resonance is difficult to characterize with simple features as it is dependent on several factors including the physio-anatomy of the speaker, the context, etc. For unvoiced sounds, hypernasal speech results in imprecise consonant production---the characteristic insufficient closure of the velopharyngeal port renders the speaker unable to build sufficient pressure in the oral cavity to properly form plosives, causing the air to instead leak out through the nose \cite{woops}. 






\subsection{Related work}

Spectral analysis of speech is a potentially effective method to analyze hypernasality.
Acoustic cues based on formant F1 and F2 amplitudes, bandwidths, and pole/zero pairs \cite{yu2009classifying}, \cite{KOZAKIYAMAGUCHI200521}, \cite{kataoka2001relationship}, \cite{hawkins}, \cite{glass}, \cite{vijayalakshmi2009selective}, \cite{pruthi}, \cite{vijayalakshmi} and changes in the voice low tone/high tone ratio \cite{lee2006voice} \cite{lee2009evaluation} have been proposed to detect or evaluate hypernasal speech. 
These spectral modifications in hypernasal speech will have an impact on articulatory dynamics, thereby affecting speech intelligibility.
Statistical signal processing methods that seek to reverse these cues, such as suppressing the nasal formant peaks and then performing peak-valley enhancement, have demonstrated improvement in the perceptual qualities of cleft palate and lip-caused hypernasal speech \cite{vikram2016spectral}, further demonstrating the connection between these cues and intelligibility.
The large variability of speech degradation patterns across neurological disease or injury challenges simple features that are based on domain expertise \cite{orozco2015characterization}.
Overall, these simple features are not robust to the complicated acoustic patterns that emerge in hypernasality, and are prone to high false positive and negative error rates in out-of-domain test cases.


In response, data-derived representations of hypernasality that combine more elemental speech features and supervised learning have been proposed. Mel-frequency cepstral coefficients (MFCCs) and other spectral transformations \cite{rah2001noninvasive}, \cite{ling}, \cite{orozco2015characterization}, \cite{rendon2011automatic}, \cite{nikitha2017hypernasality}, \cite{dubey2016zero}, \cite{dubey2018pitch}, \cite{vogel}, \cite{kataoka1996spectral}, glottal source related features (jitter and shimmer) \cite{castellanos2006acoustic}, \cite{dubey}, difference between the low-pass and band pass profile of the Teager Energy Operator (TEO) \cite{cairns1996noninvasive}, \cite{maier2008analysis}, and non-linear features \cite{orozco2012automatic}, \cite{orozco2013nonlinear} have all been proposed as model input features. Gaussian mixture models (GMM), support vector machines, and deep neural networks have been used in conjunction with these features for hypernasality evaluation from word and sentence level data \cite{nieto2014pattern}, \cite{golabbakhsh}, \cite{cairns}, \cite{ayt}. Recently, end-to-end neural networks taking MFCC frames as input and producing hypernasality assessments as output have also been proposed \cite{hypernn}. 

These methods rely on supervised learning and are trained on small data sets.  For our application they run the risk of overfitting to the data by focusing on associated disease-specific symptoms rather than the perceptual acoustic cues of hypernasality itself.

Features based on automatic speech recognition (ASR) acoustic models targeting articulatory precision have been used in nasality assessment systems \cite{maier2008analysis}. The nasal cognate distinctiveness measure uses a similar approach but assesses the degree to which specific stops sound like their co-located nasal sonorants \cite{ncd}.

 

\subsection{Contributions}



We propose a novel set of features called ``Nasalization-Articulation Precision'' (NAP), addressing the limitations of the current methods with a hybrid approach that brings together domain expertise and machine learning. Our approach relies on a combination of two minimally-supervised acoustic models trained using only healthy speech data, with dysarthric speech data only used for training simple linear classifiers on top of the features. The first model learns a distribution of acoustic patterns associated with nasalization of voiced phonemes and the second model learns a distribution of acoustic patterns of precise articulation for unvoiced phonemes. For a dysarthric sample, the model produces phoneme-specific measures of nasalization and precise articulation. As a result, the features are intuitive and interpretable, and focus on hypernasality rather than other co-modulating factors.

In contrast to other approaches that rely on machine learning, the feature models are not trained on any clinical data. We show that these features can be combined using simple linear regression to develop a robust estimator of hypernasality that generalizes across different dysarthria corpora, outperforming both neural network-based and engineered feature-based approaches.  Furthermore, we demonstrate that this estimator indeed robustly captures the perceptual attributes of hypernasality by training on our full dysarthria corpus and evaluating on a cleft lip and palate (CLP) dataset, with speakers who are, apart from the cleft palate-induced hypernasality, otherwise healthy speakers. 

To evaluate the efficacy of this new feature set, we train and validate a linear model that estimates clinician-rated hypernasality scores across different neurological diseases to ensure that it is focusing on hypernasality and not other disease-specific dysarthria symptoms. We show that this representation correlates strongly with clinical perception of hypernasality across several different neurological disorders, even when trained on one disorder and evaluated on a different disorder. Such assessment has a potential advantage as an inexpensive, non-invasive and simple-to-administer method, scalable to large and diverse populations \cite{cherney2012telerehabilitation} \cite{jacob}. 


%


\section{\label{sec:methods} Methods}

\begin{figure*}[t]
\centering
\includegraphics[width=7in]{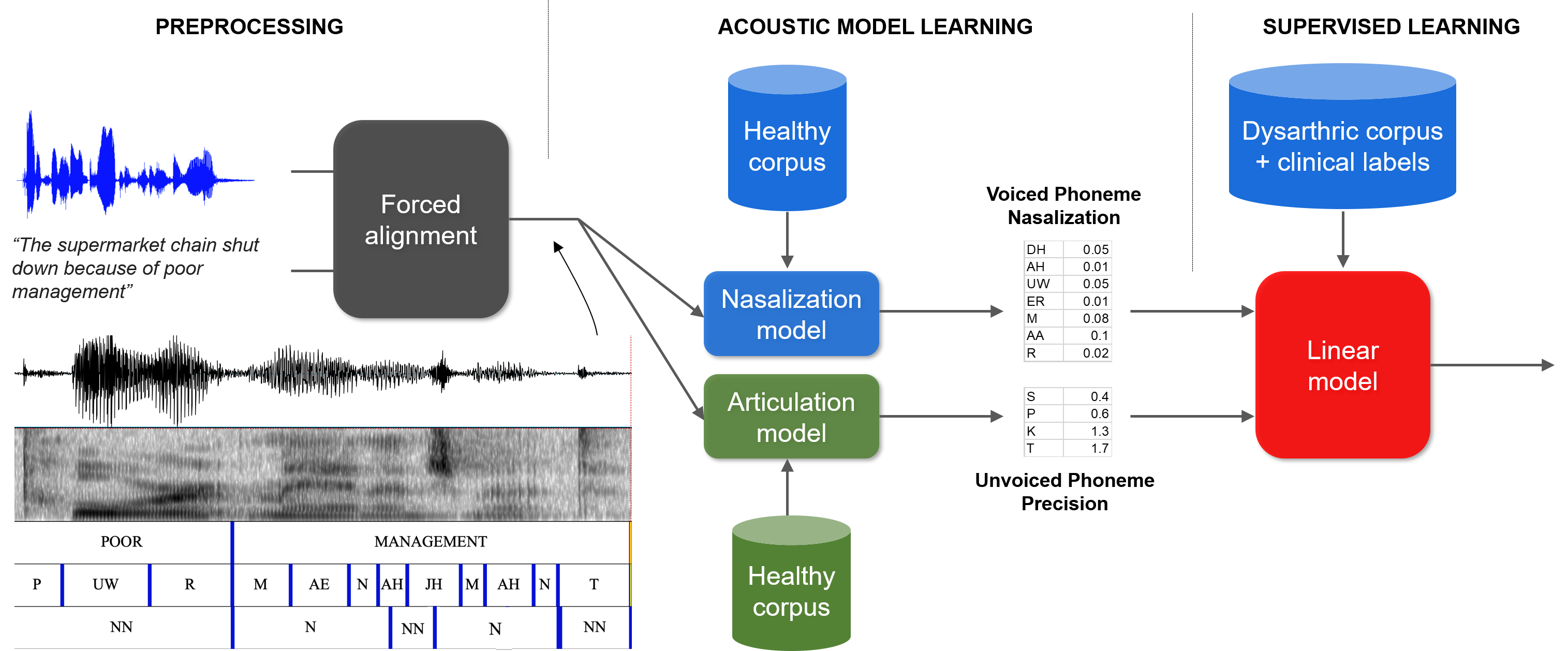}
\caption{\protect\centering A high-level diagram of the NAP method. The leftmost pre-processing segment depicts the forced alignment of transcript to audio as well as the aligned word-phoneme-nasal class segmentation of the speech signal and spectrogram.}
\label{fig:model}
\end{figure*}		

The proposed hypernasality evaluation algorithm is based on the intuition that as the severity of hypernasality increases, two broad perceptible changes take place: the unvoiced phonemes become less precise and the voiced phonemes become nasalized. To that end, we model these perceptual changes at the phone level. In Fig. \ref{fig:model}, we provide a high-level overview of the proposed hypernasality score estimation scheme. 

After forced alignment and pre-processing, the input speech is routed phoneme-by-phoneme to one of two acoustic models. The voiced phonemes are analyzed using the first model, yielding an objective estimate of acoustic nasalization based on a likelihood ratio. The unvoiced phonemes are analyzed with the second model, which captures the production quality of unvoiced phonemes, objectively estimating articulatory precision as another likelihood ratio. Both the acoustic nasalization model and the articulatory precision model are trained using healthy speech from the LibriSpeech Dataset \cite{librispeech}. The features from these models are averaged by phoneme, then input to a simple linear model to predict clinician-assessed hypernasality ratings across several different neurological disorders. We describe the data and processing steps in detail below.

\subsection{Data}

\subsubsection{Healthy speech corpus}

LibriSpeech is a public domain corpus of healthy English utterances with transcripts. It contains roughly 1000 hours of speech from 1,128 female and 1,210 male speakers reading book passages aloud, sampled at 16 kHz. It contains 100 and 360 hour ``clean'' samples, which have been carefully segmented and aligned \cite{librispeech}. We use this corpus to train both acoustic models shown in Fig. \ref{fig:model}.

\subsubsection{Dysarthric speech corpus}

The dataset contains recordings from 75 speakers (40 male and 35 female) of varying levels of hypernasality. The corpus contains data from speakers diagnosed with several different neurological disorders: 38 patients have Parkinson’s disease (PD), 6 have Huntington's disease (HD), 16 patients cerebellar Ataxia (A), and 15 patients amyotrophic lateral sclerosis (ALS).

All individuals read the same set of five sentences, capturing a range of phonemes. Reading is an ideal stimulus for this task because it controls for phonetic distributional variations that would be present in more spontaneous speech and enables for consistency between speakers and between assessments in-time, ideal qualities for a clinical measure.

The perceptual evaluation of hypernasality from recorded samples was carried out by 14 different speech language pathologists on a scale of 1 to 7.  The average hypernasality score for each speaker was used as the ground truth. The inter-rater reliability of the SLPs was moderate, with an average inter-clinician mean absolute error of 1.44 on the 7-point scale. The sentences spoken were:

\begin{enumerate}
\item The supermarket chain shut down because of poor management. 
\item Much more money must be donated to make this department succeed. 
\item In this famous coffee shop they serve the best doughnuts in town.
\item The chairman decided to pave over the shopping center garden.
\item The standards committee met this afternoon in an open meeting.
\end{enumerate}

\begin{figure}[t]
\centering
\includegraphics[width=0.9\linewidth]{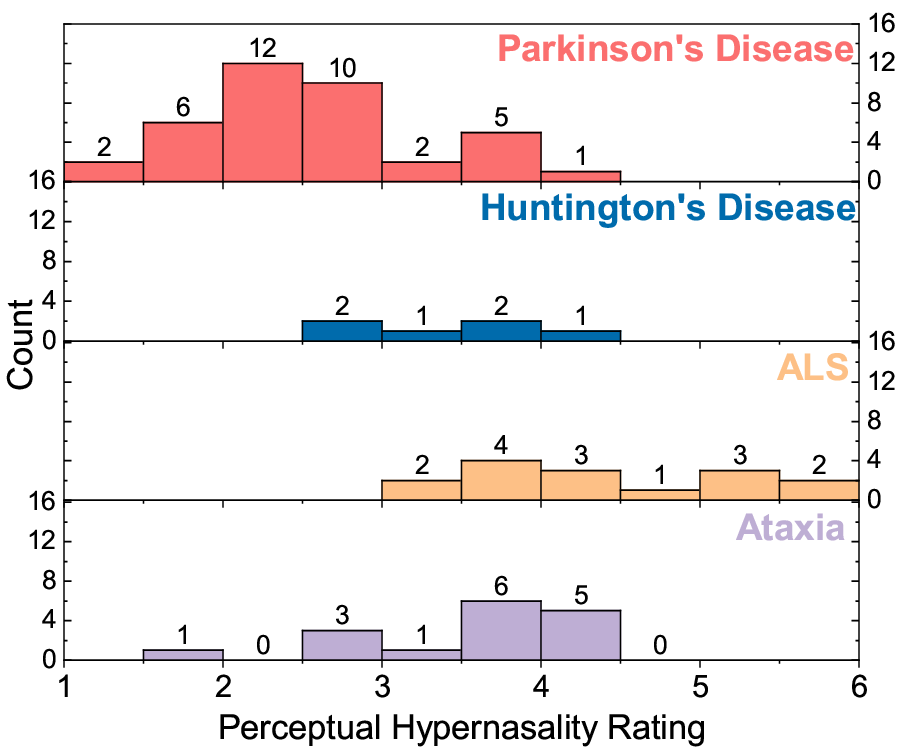}
\caption{Distribution of clinician-rated hypernasality score by disorder.}
\label{fig:histograms}
\vspace*{-5mm}
\end{figure}

The speech recordings were carried out in sound-treated room using a microphone. Table \ref{tab:table1} shows the breakdown of clinical characteristics of the subjects and the statistics of the nasality score (NS) subsets. S.D. denotes standard deviation. Figure \ref{fig:histograms} contains the clinician hypernasality score histograms for each disorder population. 

\begin{table}[h!]
\caption{\label{tab:table1}Clinical characteristics and nasality scores of the subjects. }
\scalebox{0.9}{
\begin{tabular}{ccccccc}
 Disease & Male & Female & Mean Age & S.D. Age & Mean NS & S.D. NS\\
\hline
PD & 20 & 18 & 71.06 & 9.62 & 2.55 & 0.75\\
A & 6 & 10 & 62.47 & 14.05 & 3.58 & 0.68\\
ALS & 8 & 7 & 59.54 & 13.23 & 4.41 & 0.85\\
HD & 6 & 0 & 58.40 & 13.20 & 3.31 & 0.59\\
Total & 40 & 35 & 65.80 & 12.67 & 3.20 & 1.04\\
\end{tabular}}
\vspace*{-1mm}
\end{table}

\subsubsection{Cleft Palate speech corpus}

While we are chiefly concerned with evaluating hypernasality in dysarthric speakers exhibiting neuromuscular diseases, cleft lip and palate (CLP) speech is useful for evaluation purposes, as CLP speech often exhibits hypernasality without the other kinds of perceptual changes (slurring, generalized articulatory imprecision) that also arise in dysarthria. We use a corpus of 6 child and 12 adult CLP speakers with different levels of hypernasality severity, that span the hypernasality range (from normal to extreme) in equal intervals \cite{kuehn2002efficacy} to demonstrate that our model chiefly captures hypernasality rather than any associated neurologically disordered speech symptoms. 


\subsection{\label{subsec:3:0} Data pre-processing}


Consider an utterance $x(t)$ with sampling rate $F_s$ and a corresponding transcript of phonemes $p_j$, $\{p_1, p_2, \dots p_{N_p}\}$. We analyze $x(t)$ with a $20\mathrm{ms}$ frame length and $10\mathrm{ms}$ overlap. For a frame indexed by $i$, $x_i(t)$, we extract a set of features, $\mathbf{x}_i$. The utterance $x(t)$ is force-aligned using the Montreal Forced Aligner\footnote{See Section \ref{subsec:fa} for discussion on forced alignment performance for these dysarthric speech samples.} \cite{mfa} at the phoneme level. We denote the data feature matrix for all frames that are aligned to phoneme $p_j$ by $\mathbf{X}^{p_j}$. 

After alignment, we use different features for the acoustic nasalization model than for the articulatory precision model. For the nasalization model we use perceptual linear prediction (PLP) features because they better preserve acoustic cues that have been previously used to model hypernasality, including formant frequencies, bandwidths, and spectral tilt \cite{hermansky1990perceptual}.  For the articulatory precision model for unvoiced phonemes, we use mel-frequency cepstral coefficients (MFCCs), a common representation for automatic speech recognition applications \cite{muda2010voice}. We also use a lower sampling rate for the voiced nasalization model (8 kHz) compared to the unvoiced articulatory precision model (16 kHz). This is motivated by the difference in spectral energy distribution between voiced and unvoiced sounds.

\subsection{\label{subsec:3:1}  Nasalization model}

To assess nasalization on a phoneme level, we model the distributions of two classes of voiced phonemes. The ``oral'' non-nasal ($ORL$) class consists of all voiced oral consonants and all vowels from syllables where nasal consonants are not present. Similarly, we define the ``nasal'' class ($NAS$) to contain the nasal consonants as well as half of adjacent vowels surrounding them. These rules were implemented after alignment; an illustrative example of the two classes is shown in the third tier of the aligned example in Fig. \ref{fig:model}.

For this task, we use 100 hours of clean-labeled speech from the LibriSpeech dataset \cite{librispeech}. We first perform forced phone-alignment to the transcript as shown in Figure \ref{fig:model}. We partition all phonemes into the $NAS$ and $ORL$ classes. For each frame in each phoneme, we extract 13 PLP coefficients, giving two feature matrices, $\mathbf{X}^{NAS}$ and $\mathbf{X}^{ORL}$, containing all frames of nasal PLPs in one, and non-nasal PLPs in the other. To model the probability density functions, we use a 16-mixture Gaussian Mixture Model (GMM). The weight, mean, and covariance matrix for each of the GMM components is learned using the expectation maximization (EM) algorithm. The GMM for the nasal class is represented by $\lambda_{NAS}=\{\mu_{NAS}, \Sigma_{NAS}, \omega_{NAS}\}$, $i = 1, 2, ... 16$. Here, $\mu_{NAS}$, $\Sigma_{NAS}$ and $\omega_{NAS}$ represent the mean, covariance matrix and weight of the $i^\textrm{th}$ Gaussian, respectively. Similarly, for the non-nasal class the GMM components are given by $\lambda_{ORL}=\{\mu_{ORL}, \Sigma_{ORL}, \omega_{ORL}\}$, $i = 1, 2, ... 16$.

After training on healthy speech, we provide a segmented dysarthric utterance to evaluate the likelihood from each of the two learned probability density functions. For an out-of-sample input, we estimate the likelihood, voiced phoneme by voiced phoneme. That is, for data feature matrix $\mathbf{X}^{p_j}$, the likelihood that this phoneme is nasalized is
\begin{equation}
  f(\mathbf{X}^{p_j} | \lambda_{NAS}) = \prod_{i \in p_j} f(\mathbf{x}_{i} | \lambda_{NAS}), 
\end{equation}
where the notation $i \in p_j$ is shorthand notation for all $20\mathrm{ms}$ frames aligned to phoneme $p_j$. Similarly for the $ORL$ class, 
\begin{equation}
  f(\mathbf{X}^{p_j} | \lambda_{ORL}) = \prod_{i \in p_j} f(\mathbf{x}_{i} | \lambda_{ORL}).
\end{equation}






We use the log-likelihood ratio test statistic as a continuous measure of nasalization. In particular, we define
\begin{equation}
N(p_j) = \mathrm{log} \left ( \frac{   f(\mathbf{X}^{p_j} | \lambda_{NAS}) }{  f(\mathbf{X}^{p_j} | \lambda_{ORL}) } \right )/|\mathbf{X}^{p_j} |,
\end{equation}
where $|\mathbf{X}^{p_j} |$ represents the number of acoustic frames aligned to phoneme $p_j$. This statistic is calculated for every voiced, non-nasal phoneme in the input utterance. Thus, for a given speaker, a nasalization log-likelihood ratio is computed for all the voiced phonemes, 
(AA, AE, AH, AO, AW, AY, B, D, DH, EH, ER, EY, G, IY, JH, V, Z)\footnote{In this work we use ARPAbet codes to transcribe English phonemes \cite{zue1996transcription}.}. 

\begin{figure}[t]
\centering
\includegraphics[width=0.85\linewidth]{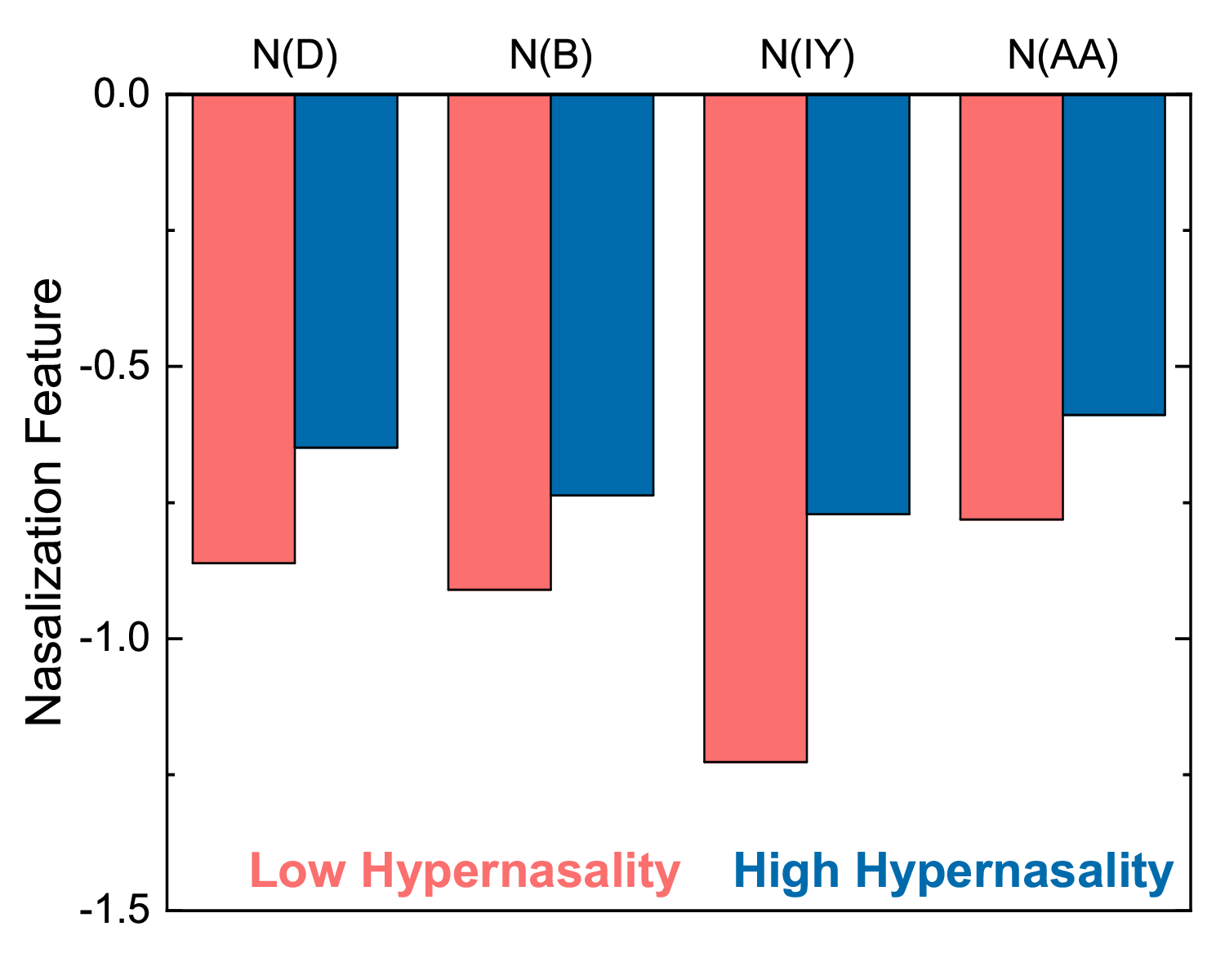}
\caption{Bar chart of the nasalization feature for D, B, IY, and AA for low hypernasality and high hypernasality speakers.}
\label{fig:nfseq}
\end{figure}

For non-nasalized speech, we expect the value of $N(\mathbf{X}^{p_j})$ to be low, whereas for nasalized speech, we expect it to be high. Figure \ref{fig:nfseq} shows a comparison of the value of the nasalization likelihood feature between a group of high hypernasality ($>4$ perceptual rating) and a group of low hypernasality ($<3$ Perceptual rating). We average the hypernasality scores for the 4 most relevant phonemes for predicting hypernasality (see Section III-B for details). As expected, there is an increase in the nasality feature value for an increase in severity of hypernasality.

\subsection{\label{subsec:3:2}  Articulation model}

The articulation model is an implementation of the goodness of pronunciation algorithm \cite{wittPp} based on an acoustic model trained using Kaldi as specified in \cite{ming}. For our implementation we used a triphone model trained with a Gaussian Mixture Model-Hidden Markov Model on 960 hours of healthy native English speech data from the LibriSpeech corpus \cite{librispeech}. The input features to the ASR model are a $39$-dimensional second-order Mel-Frequency Cepstral Coefficient (MFCC) (including deltas and delta-deltas) with utterance-level cepstral mean variance normalization and Linear Discriminant Analysis transformation. We use the Kaldi toolkit training scripts for training the model.  In contrast to the nasalization model, here we use a sampling rate of 16 kHz to capture the wideband nature of unvoiced phonemes. The nasalization model required a much smaller training set (100 hours) since there were only two classes modeled by the GMM.

After training, the acoustic model can be queried using the Viterbi decoding algorithm for the posterior probability $P(\mathbf{X}|q)$ of a given set of acoustic feature frames $\mathbf{X}$ representing a realization of some phoneme $q$. For a ``well-articulated'' phoneme, no phoneme apart from the one intended by the speaker should maximize this posterior.

We use the acoustic model to assess articulatory precision as follows. Considering the set of phonemes $Q$ in the language, we assess the log-likelihood ratio of the frames $\mathbf{X}^{p_j}$ from a given phoneme $p_j$, to the maximum log-likelihood across all phonemes, 
\begin{equation}
AP(p_j)=\log\Big(\frac{P( \mathbf{X}^{p_j} |p_j)}{\mathrm{max}_{q\in Q}P( \mathbf{X}^{p_j} |q)}\Big)/|\mathbf{X}^{p_j} |,
\end{equation}
where $|\mathbf{X}^{p_j} |$ represents the number of acoustic frames aligned to phoneme $p_j$.

This processing is performed after forced alignment to the transcript labels, and assessed for each unvoiced phoneme to permit by-phoneme analysis of precise articulation. 

\begin{figure}[t]
\centering
\includegraphics[width=0.85\linewidth]{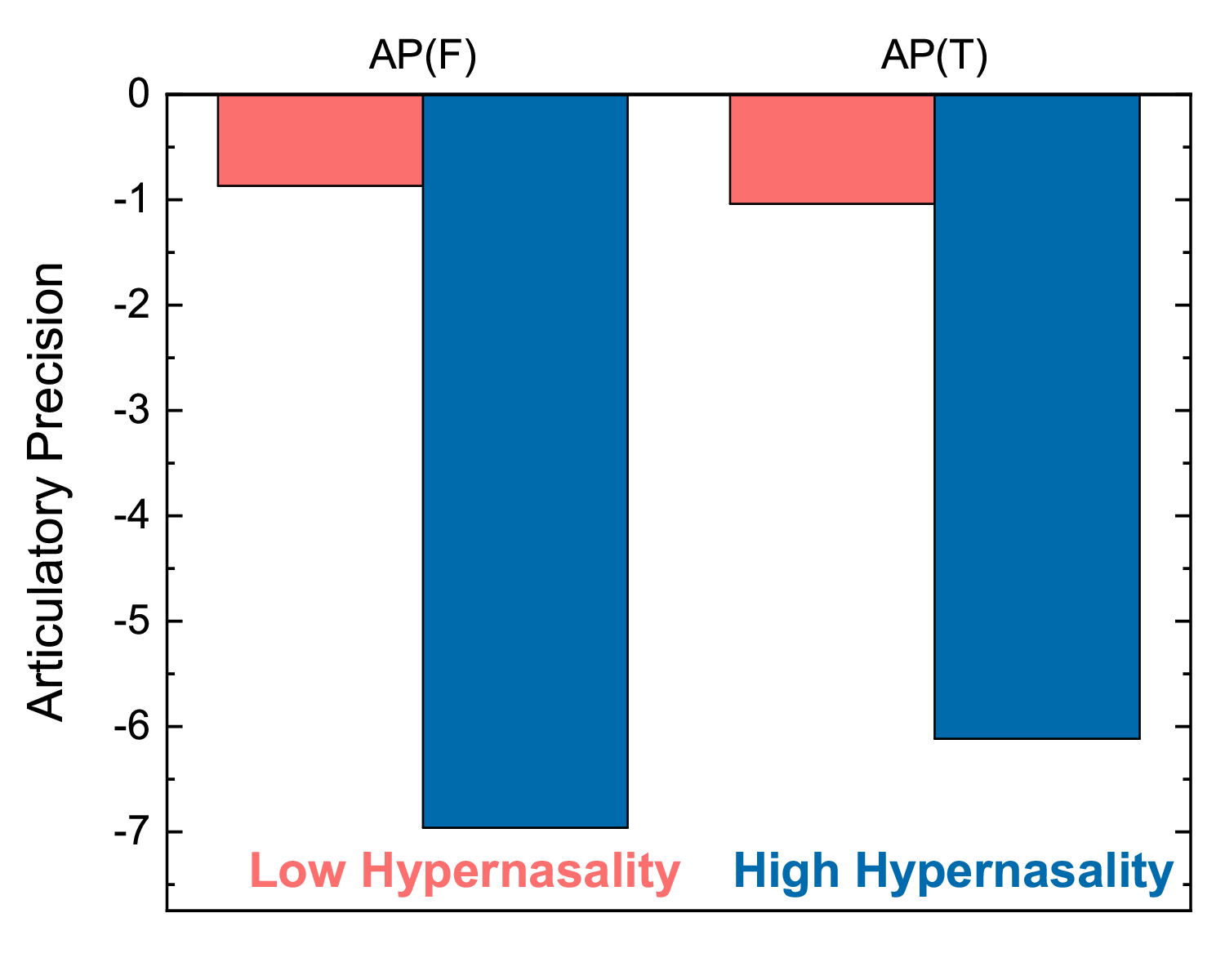}
\caption{Bar chart of the articulatory precision feature for F and T for low hypernasality and high hypernasality speakers.}
\label{fig:gopseq}
\end{figure}	

For speakers who exhibit little hypernasality, we expect the value of $AP(\mathbf{X}^{p_j})$ for unvoiced phonemes to be high, whereas for hypernasal speakers, we expect it to be lower. Figure \ref{fig:gopseq} shows a comparison of the average value of the articulation precision feature between a group of high hypernasality ($>4$ perceptual rating) subjects and a group of low hypernasality ($<3$ Perceptual rating) subjects. We average the articulation scores for the most relevant phonemes for predicting hypernasality (see Section III-B). As expected, there is a decrease in the articulation precision feature value for an increase in severity of hypernasality. Furthermore, we expect hypernasality to exhibit unique patterns in terms of affected and unaffected unvoiced phonemes, which are not general to dysarthria \cite{ncd}, making phoneme-level AP classification a valuable signal in quantifying hypernasality.

\begin{table*}[h]
\label{table:nasest}
  \centering
    \caption{Comparative evaluation of state-of-the-art formant features (FF-Linear, -Additive, -KNN) and neural network-based (MFCC-, PASE-NN) approaches and the clinician raters (Human) against our NAP features for predicting clinician hypernasality score, conducted using leave-one-speaker-out (LOSO) and using leave-one-disease-out (LODO) cross validation. We report mean absolute error (MAE) on the 7-point scale and Pearson correlation coefficient (PCC). Bold denotes \textbf{best performance overall} for a metric, italic denotes \textit{best non-human performance} for a metric when applicable.}
 \begin{tabular}{|p{2cm}||p{1cm}|p{1cm}|p{1cm}|p{1cm}|p{1cm}|p{1cm}|p{1cm}|p{1cm}|p{1cm}|p{1cm}|}
 \hline
 Train on & \multicolumn{2}{c|}{Ataxia, HD, PD, ALS} & \multicolumn{2}{c|}{HD, PD, ALS}  & \multicolumn{2}{c|}{Ataxia, PD, ALS} & \multicolumn{2}{c|}{Ataxia, HD, ALS} & \multicolumn{2}{c|}{Ataxia, PD, HD} \\
 \hline 
 Test on & \multicolumn{2}{c|}{Left-out speaker} & \multicolumn{2}{c|}{Ataxia}  & \multicolumn{2}{c|}{HD} & \multicolumn{2}{c|}{PD} & \multicolumn{2}{c|}{ALS} \\
 \hline
 Model  & MAE    & PCC &   MAE & PCC & MAE & PCC    & MAE &   PCC & MAE & PCC \\
\hline
FF-Linear &   0.871  &  0.180   & 0.823 & 0.042  & 0.666  &   -0.751   &  1.316   & 0.351 & 1.426 & -0.425     \\
FF-Additive &   0.789  &  0.435   & 0.730 & -0.123  & 0.693  &   -0.557   &  1.334   & 0.277 & 1.260 & 0.429     \\
FF-KNN &   0.754  &  0.481   & 0.781 & 0.333  & 0.567  &   0.381   &  1.218   & 0.402 & 1.227 & -0.039     \\
MFCC-NN   & 0.884 & 0.458 & 0.904    & -0.120 &   \textbf{0.429} & 0.568 & 0.800 & 0.457    & 1.233 &   0.315\\
PASE-NN & 0.774 & 0.417 & 0.707 & -0.204  & 0.433  & 0.237  & 1.150 & 0.163 &  1.407 & 0.176  \\
NAP-Linear \textit{ours} &   \textbf{0.587} &  \textit{0.722}  & \textbf{0.546} & \textbf{0.750} & 0.559 &   \textbf{0.737}  & \textbf{0.509}   & \textbf{0.697} & \textbf{0.597} & \textit{0.527} \\
\hline
Human & 0.832 & \textbf{0.725} & 0.871 & 0.476 & 1.256 & 0.550 & 0.746 & 0.636 & 0.979 & \textbf{0.601} \\
\hline
\end{tabular}
\end{table*}

\begin{figure*}[ht]
\begin{minipage}{.33\linewidth}
\centering
\label{main:a}
\subfloat[]{\includegraphics[width=1\linewidth]{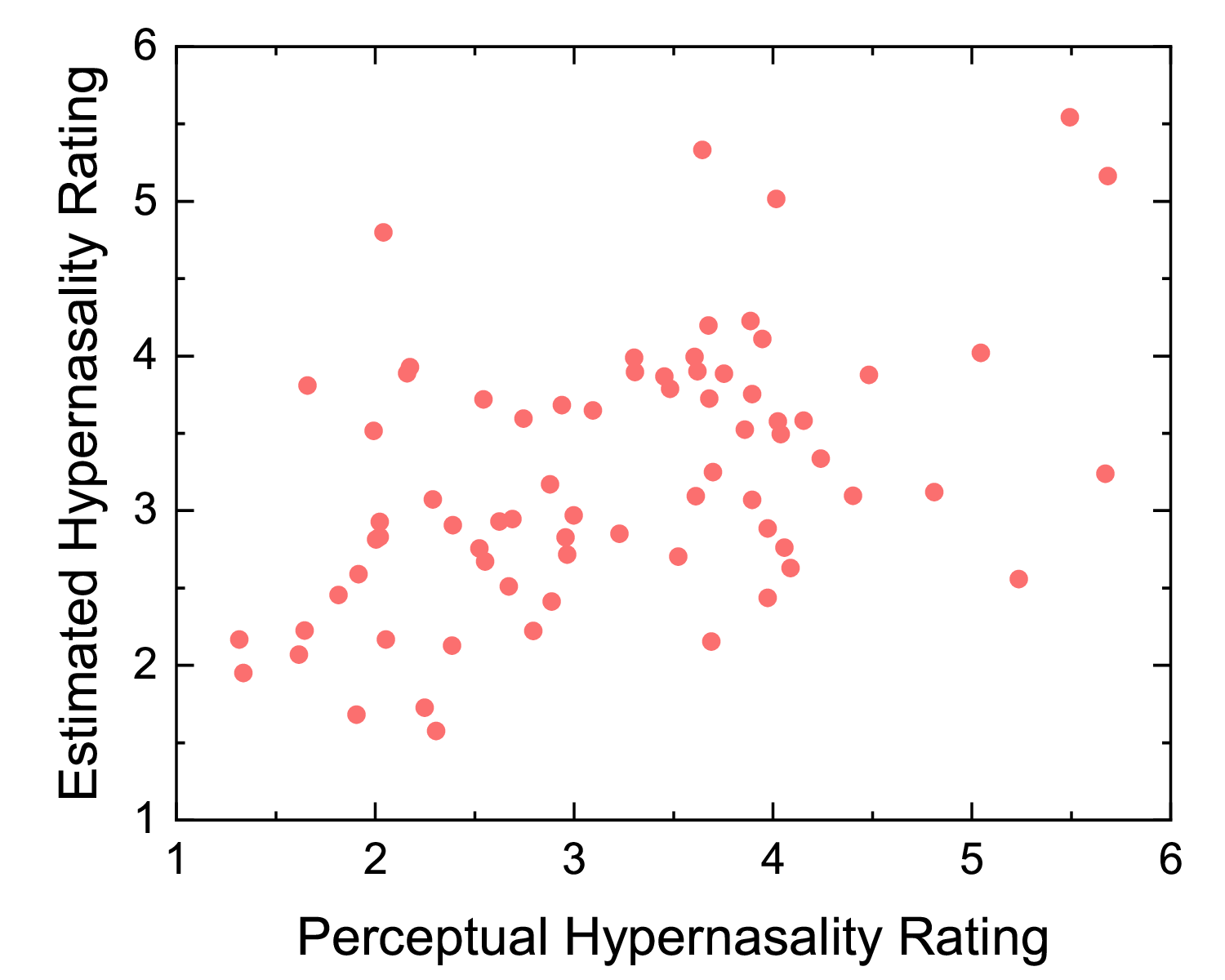}}
\end{minipage}%
\begin{minipage}{.33\linewidth}
\centering
\label{main:b}
\subfloat[]{\includegraphics[width=1\linewidth]{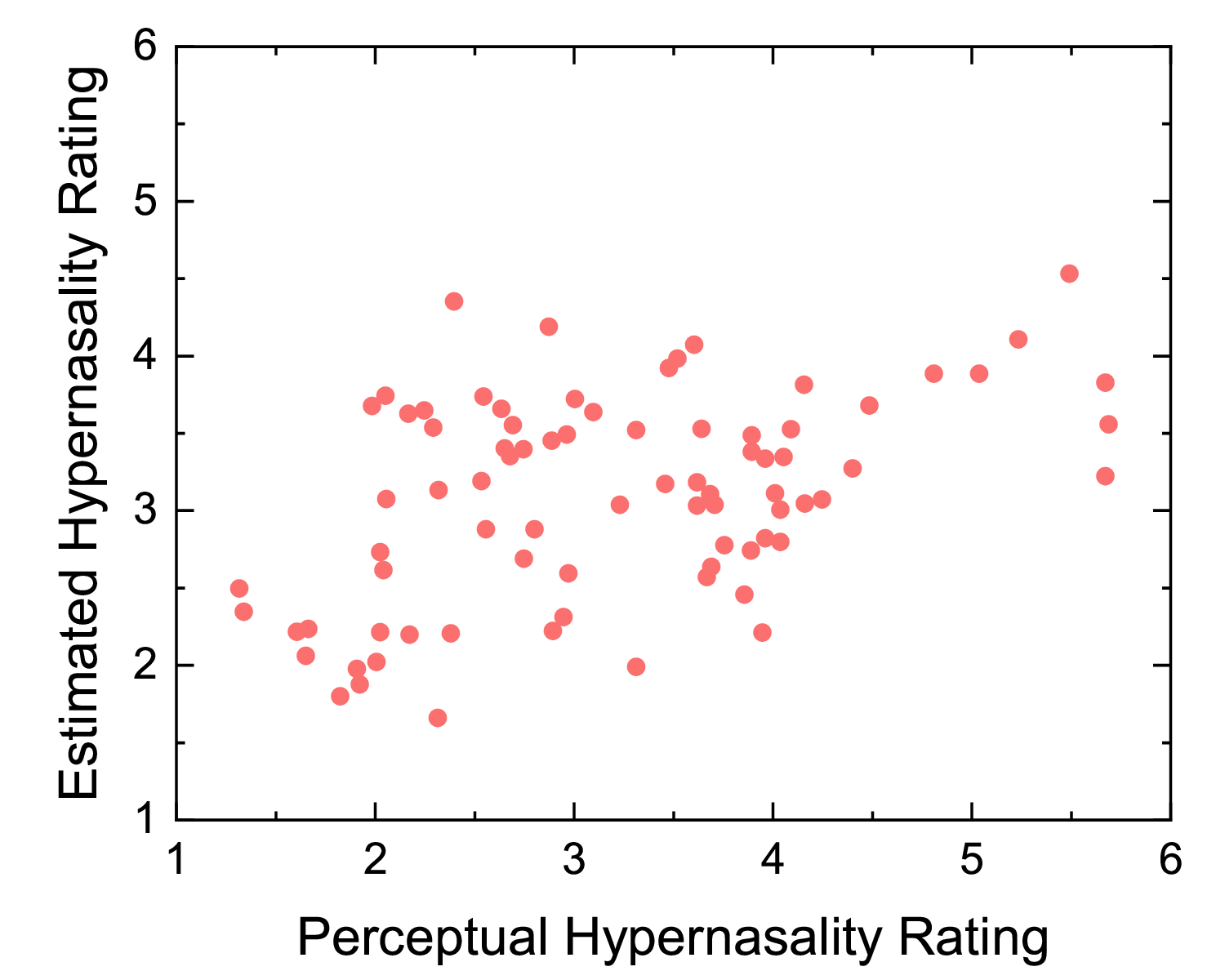}}
\end{minipage}%
\begin{minipage}{.33\linewidth}
\centering
\label{main:c}
\subfloat[]{\includegraphics[width=1\linewidth]{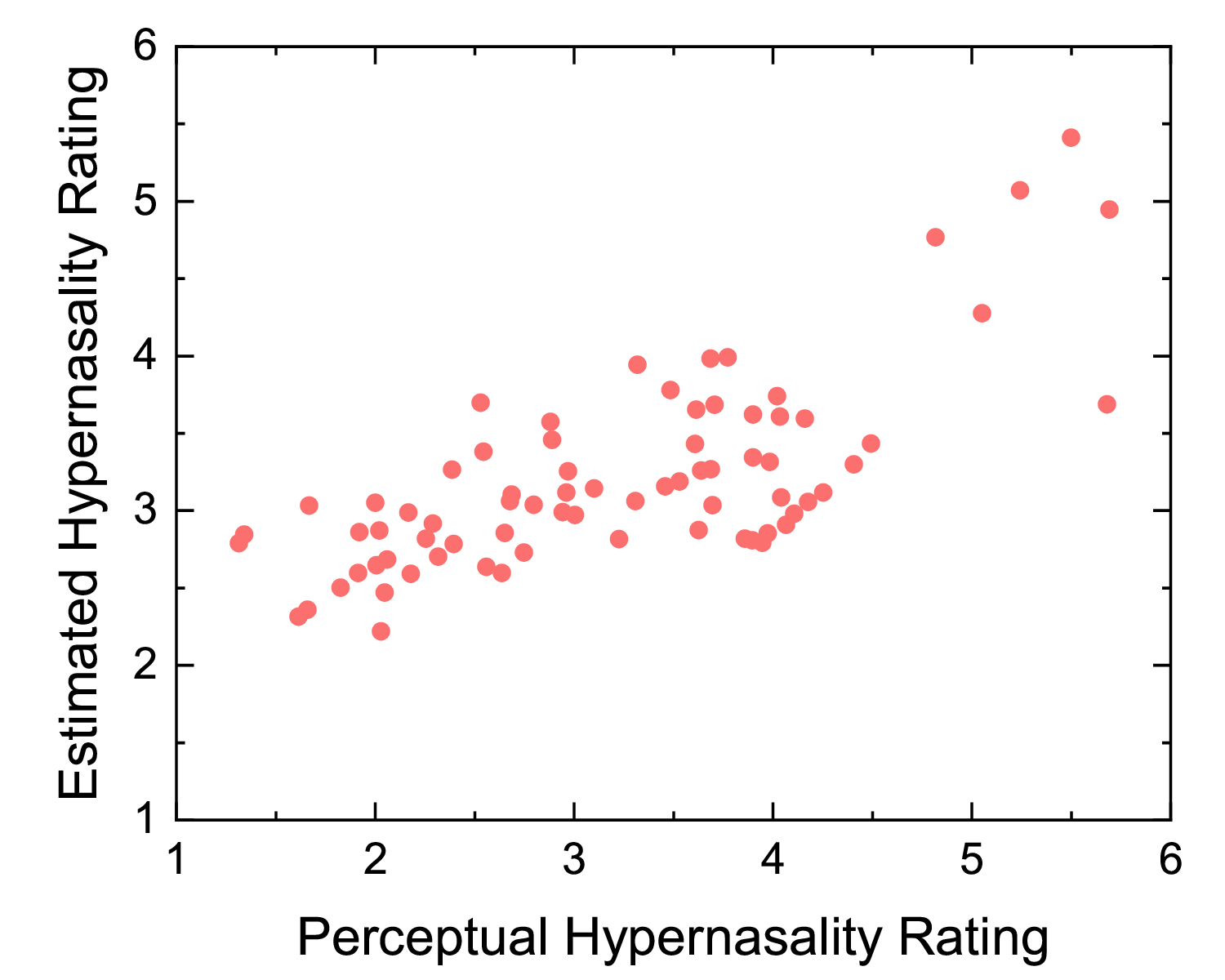}}
\end{minipage}%

\caption{LOSO results from predicting the hypernasality score for the simple feature baseline with (a) the KNN classifier and simple formant features (FF-KNN), (b) the neural network baseline (MFCC-NN), and (c) the NAP features with simple linear regression (NAP-Linear)}
\label{fig:comparisonplot}
\end{figure*}

\subsection{\label{subsec:3:3} Linear regression model}

In the interest of generalization and clinical interpretability, simple linear ridge regression models \cite{sklearn} are used to estimate the nasality score using the phoneme-averaged nasalization and articulatory precision features as input.

Two different cross-validation strategies are used to evaluate model performance and the quality of the input features. First, we use leave-one-speaker-out (LOSO) cross-validation, as is typically done. To evaluate the generalization across out-of-domain diseases, we also perform leave-one-disease-out (LODO) cross-validation. In LODO, data for three of the neurological conditions is used for training and the fourth is used for testing.

\section{Experiments and Results}

In this section we evaluate the efficacy of the features in two ways: by analyzing the correlation between individual per-feature averages and speaker hypernasality rating, and by observing the performance of a simple linear regression model directly calculating the hypernasality score for a speaker from their features. For the hypernasality score computation task, we evaluate the performance of a model against two baselines. Models are compared using mean average error (MAE) and the Pearson correlation coefficient (PCC) between the clinical perceptual hypernasality score and the predicted hypernasality scores.



\begin{figure*}[ht]
\centering
\includegraphics[width=0.95\linewidth]{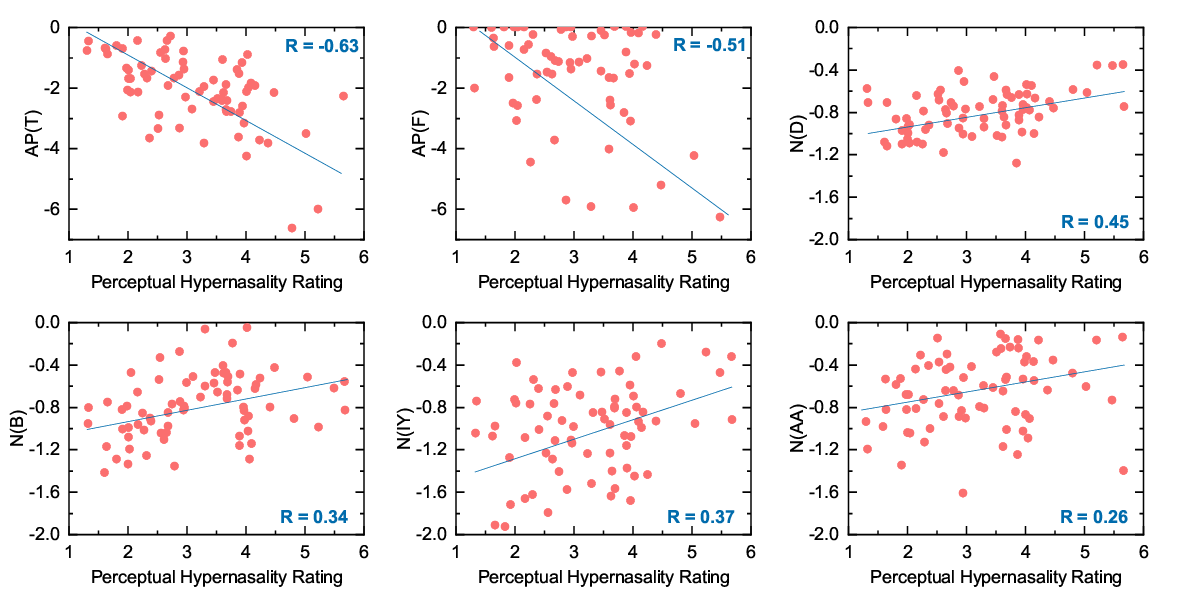}
\caption{Plots of the two most prominent articulatory precision features (AP(T) and AP(F)) as well as four of the most prominent nasalization features (N(D), N(B), N(IY), and N(AA)) against clinician-assessed nasality score.}
\label{fig:apf}
\end{figure*}

\subsection{Hypernasality evaluation}


In Table II, we show the results of the evaluations for six different models alongside comparable evaluations of human clinician raters. We evaluated our NAP features against comparison models representing the state of the art in engineered features and in supervised learning. The most predictive acoustic features for hypernasality presented in \cite{styler} were extracted using Praat source code provided by the authors \cite{styler2013using}. These formant features (FF) included $F1$ formant amplitude, $P0$ nasality peak amplitude, and normalized and raw $A1-P0$ difference \cite{chen1997}. All features were extracted for each vowel and used in a linear and non-linear model to estimate the clinician-assessed hypernasality labels. The linear model is based on simple multiple regression whereas the non-linear models are based on additive regression and $k$-nearest neighbor regression. The results of this model are labeled FF-Linear, FF-Additive, and FF-KNN in Table II. 

In addition, we evaluated two neural networks. We implemented the neural network proposed in \cite{hypernn}, consisting of three feed-forward layers with sigmoid activations. The inputs to the networks are 39-dimensional Mel Frequency Cepstral Coefficients (MFCC) computed with a 20-ms window length and no overlap. The hidden layer is of size  100, and the output layer of size 1. The output value is averaged across all frames to provide a single nasality score estimation per speaker. We also implemented a basic model using the Problem Agnostic Speech Encoder (PASE) \cite{pase}. The PASE encodings were first extracted from the raw audio, then fed through three feed-forward layers with ReLU activations, followed by a single LSTM layer, all with hidden size 250. After max-pooling in time, a final feed-forward layer projects the latent codes to the final hypernasality score estimation. Both models are trained using L1 loss and the Adam optimizer \cite{adam} for 50 epochs with a learning rate of 0.001. These models' results are reported in Table II as MFCC-NN and PASE-NN, respectively.

Finally, we treat each of the 14 clinician hypernasality severity evaluators as an individual estimator, with which we assess MAE and PCC from the ground-truth average hypernasality severity scores. For the LOSO evaluation we average the 14 human evaluator MAE and PCC scores across the 75 speakers, and then average these across the 14 evaluators to get an average human baseline MAE and PCC. Similarly, for the LODO conditions we evaluate only the evaluation disease subset. These results are presented in Table II as ``Human.''

The results show that the linear model based on NAP features not only consistently outperforms the baseline FF and NN models, achieving the best performance of all systems on all measures but LODO HD MAE. The improved performance of our system over the baselines is most pronounced on the LODO conditions, suggesting that our model is more consistently robust to disease-specific manifestations. 

The differences are also apparent when we analyze the individual LOSO correlation plots in Fig. \ref{fig:comparisonplot}. These scatter plots relate the estimated hypernasality score for each speaker against the actual hypernasality score.  As is clear from the figures, the correlation of the baseline methods is largely driven by the samples with very high nasality scores. The NAP model exhibits a linear trend between the predicted and actual values throughout the hypernasality range. 

Furthermore, our model outperforms the human evaluators in all but two measures, LOSO PCC and LODO-ALS PCC.

\subsection{Individual feature contributions}

\begin{figure}[t]
\centering
    \includegraphics[width=\linewidth]{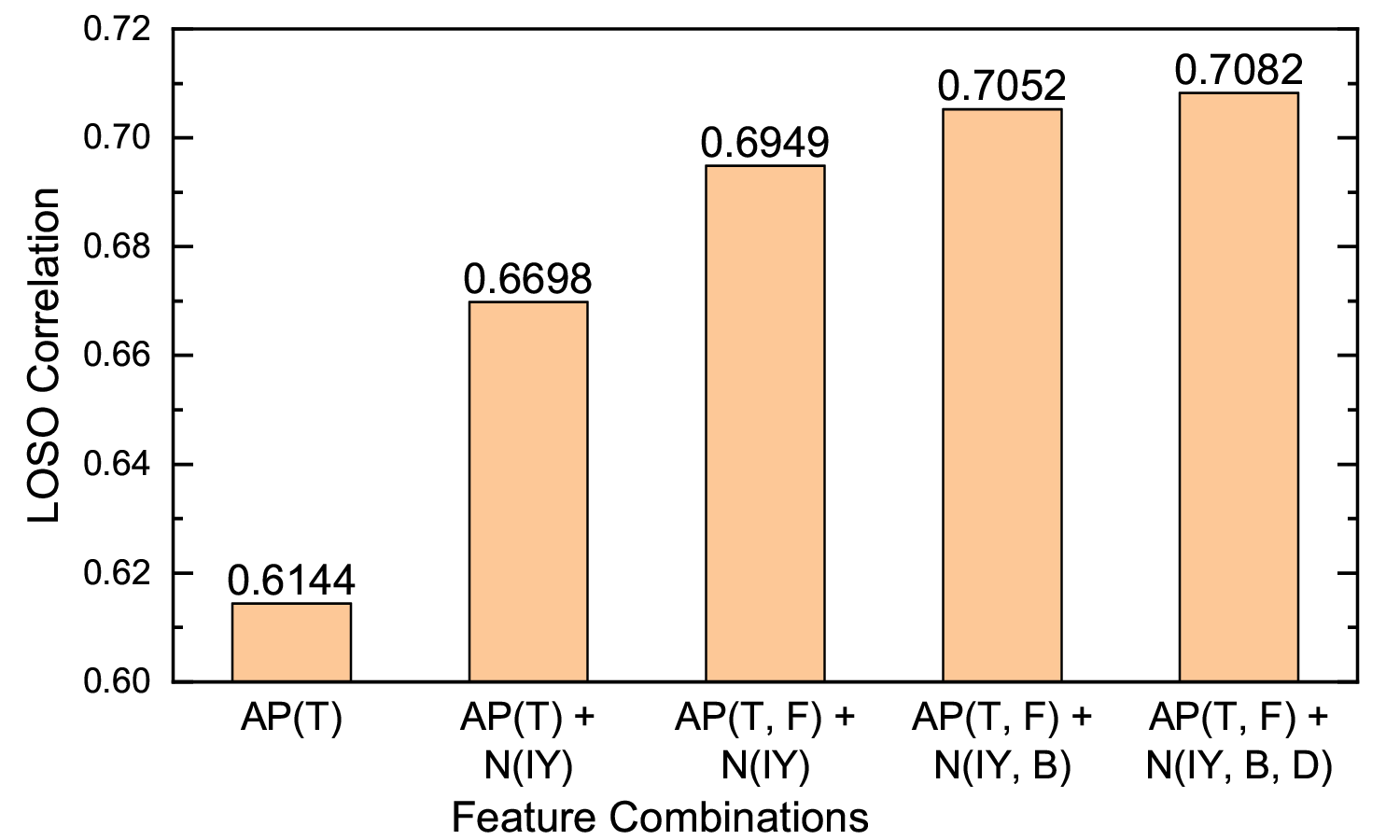}
\caption{Cumulative marginal improvement plot of leave-one-speaker-out correlation with the addition of the most optimal articulatory precision and nasalization features.}
\label{fig:cumfut}
\end{figure}

We use a simple forward selection algorithm for the LOSO model to identify the most predictive NAP features. The algorithm identifies the subset of features that minimizes the cross-validation mean square error between the model predicted and clinical hypernasality rating, with features iteratively added until the error stops decreasings. This procedure results in 6 non-redundant features selected for prediction. This includes the articulatory precision for T and F and the nasalization for D, B, IY, and AA. We plot the top features against the clinical perceptual nasality ratings in Fig. \ref{fig:apf}; Fig. \ref{fig:cumfut} depicts the marginal improvement in LOSO correlation as features are added in by decreasing feature prominence.



\section{Supplementary Analysis}

The previous section experimentally addresses the core question of whether our novel model achieves improved results in estimating hypernasality in individuals with neuromuscular disease in ways that generalize across diseases. In this section, we perform supplementary experiments to interrogate specific aspects of the features, model, and approach.

\subsection{\label{subsec:role}Role of articulatory precision in hypernasality}

Articulatory precision and hypernasality are tightly linked. Neurological VPD, which gives rise to hypernasality, is often accompanied by impaired articulatory precision. The neurological conditions we study herein impact several aspects of speech production, including respiration, voicing, resonance, and articulation. This brings up two important questions:
\begin{itemize}
    \item Do our features capture changes related to hypernasality that go beyond changes in articulatory precision?
    \item Are our features sensitive to changes in articulatory precision that result from only hypernasality (and not other articulatory impairments resulting from dysarthria)?
\end{itemize}

In an attempt to decouple articulatory precision from hypernasality, we collect clinical articulatory precision ratings (in addition to the hypernasality ratings) from the same clinicians. The inter-rater reliability of the ratings was robust, with an average pairwise Pearson correlation coefficient of 0.75 and a mean absolute error of 1.01 on a 7-point scale.

\begin{figure}[t]
\centering
    \includegraphics[width=\linewidth]{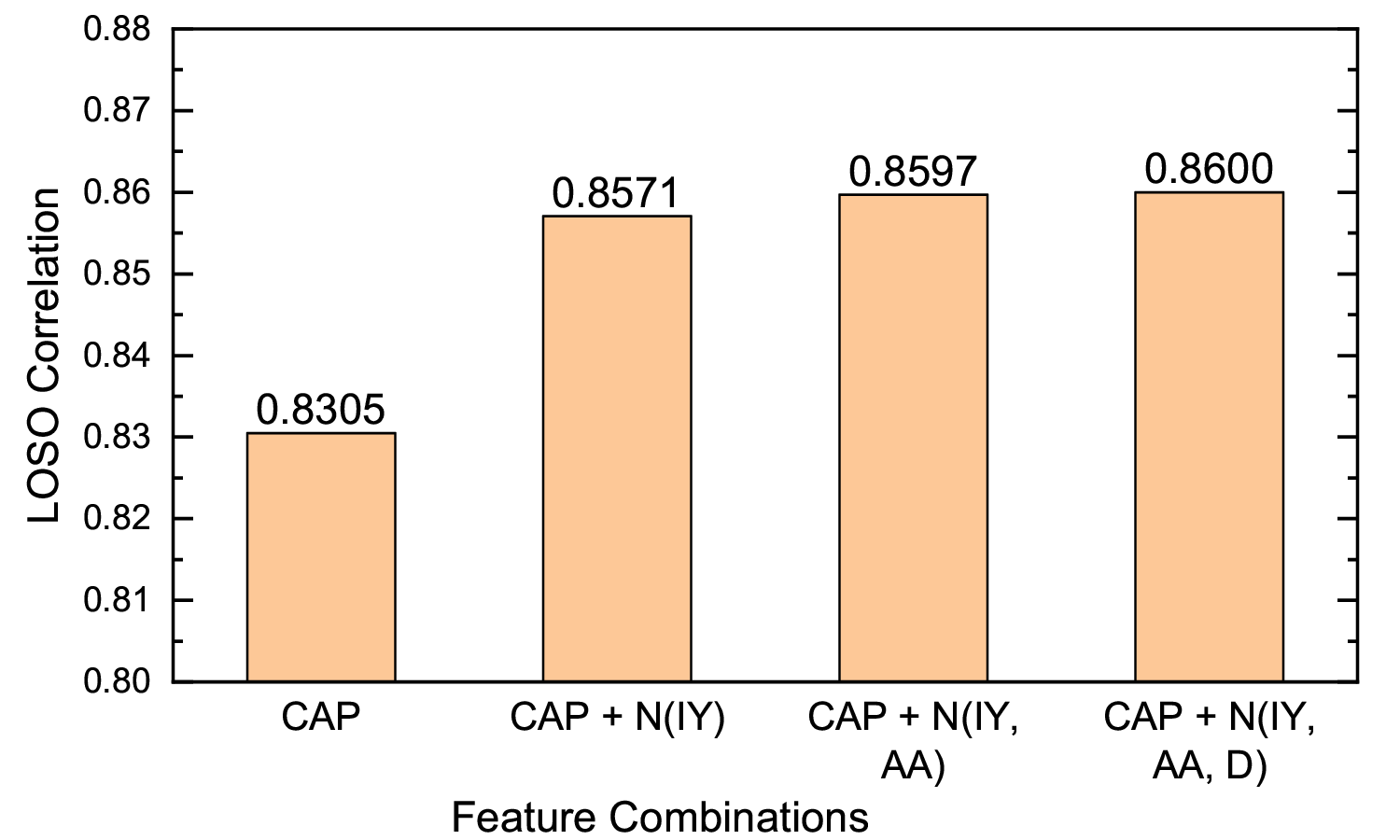}
\caption{Cumulative marginal improvement plot of LOSO correlation for the most optimal articulatory precision and nasalization features, and clinician articulatory precision (CAP).}
\label{fig:inclAP}
\end{figure}

To answer the first question above, and demonstrate that our features capture information beyond changes in articulatory precision, we use a multiple linear regression model with  clinician-rated articulatory precision alongside our six most predictive features (N(AA), N(IY), N(B), N(D), AP(T), AP(F)) as independent variables. The dependent variable is the clinical hypernasality rating. We once again use the forward selection algorithm on PCC to cumulatively select the most predictive features. The results are depicted in Figure \ref{fig:inclAP}. As expected, the subjective AP rating is most predictive as there is significant overlap with hypernasality, and it is selected first. In the presence of this generalized measure of articulatory precision, it makes sense that AP(T, F), features that are themselves estimating AP, would not be selected. This reinforces the rationale for their inclusion in the model. Three nasalization features, N(IY, AA, D), are able to further improve the correlation of the linear model predictions.



To answer the second question, and demonstrate that our features are sensitive to hypernasality alone, we evaluate a linear model trained on our full dataset of dysarthric speech using the six most predictive features predicting hypernasality scores for the 18 speech samples from individuals with cleft lip and palate in our CLP dataset. The linear hypernasality model trained on our dysarthric speech corpus achieves a hpyernasality severity predicition PCC of 0.89 for the adults and 0.82 for the children. This provides additional evidence that our features capture the perceptual quality of hypernasality and not other co-modulating symptoms.

\subsection{\label{subsec:fa}Effectiveness of forced alignment}

The features we have proposed herein rely on force aligning known transcripts to dysarthric speech \cite{yeung2015improving}. This can be problematic as co-articulation, blending, missed targets, distorted vowels, and poor articulation present in severely disordered speech \cite{green2003automatic} may interfere with the appropriate matching of dictionary phoneme-word pairs to the realized sounds \cite{foraliiss}. 

\begin{figure}[t]
\centering
\includegraphics[width=0.93\linewidth]{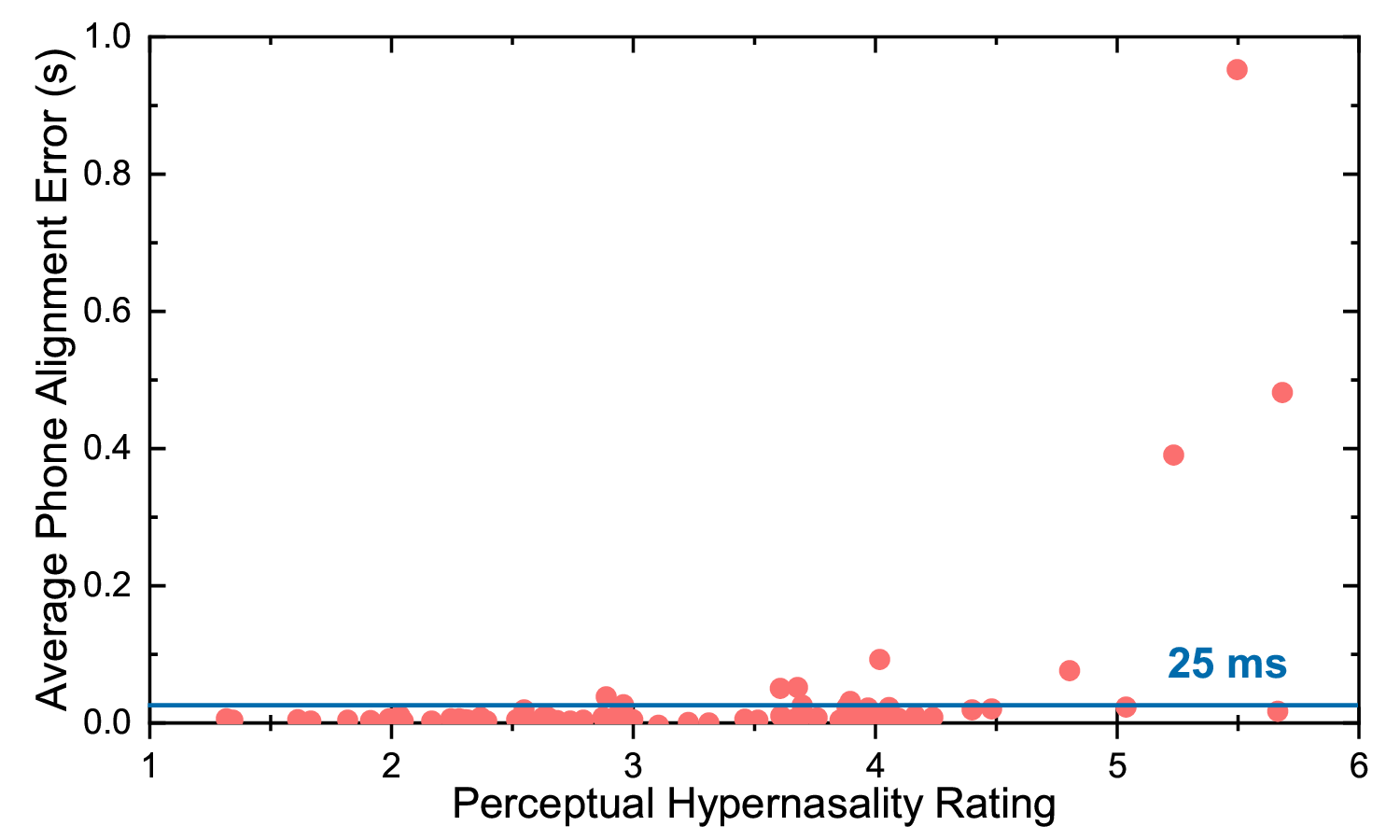}
\caption{Plot of average alignment errors per speaker (s) against clinician-rated articulatory precision at the phone level. Dashed line indicates an average alignment error of 25 ms.}
\label{fig:alierrs}
\end{figure}

We directly evaluate the prevalence of alignment errors generated by our forced alignment methodology using manually aligned transcripts. Two  annotators produced word- and syllable-level aligned transcripts using the same spelling and phoneme-word conventions employed in the acoustic model dictionary for all utterances in the dataset. For each speaker we count word- and phone-level alignment errors based on the position of the center point of a word or phoneme $t_c'$ as assessed by the forced aligner and the beginning and end of the corresponding word or syllable, $t_{min}$, $t_{max}$ as assessed by the human transcriber. For each word or phoneme, the error is counted as $t_e = \max(0,t_{min}-t_c',t_c'-t_{max})$. This error measure returns 0 if the center of the phoneme falls within the syllable; otherwise it returns the maximum error between the center of the automatically aligned phoneme and the start and end of the manually-aligned syllable. In Figure \ref{fig:alierrs} we show the alignment error (in sec.) against the hypernasality rating to show how alignment error rates progress as hypernasality increases. The results show that for all but the most severely hypernasal speakers forced alignment works effectively.

These results also indicate that our objective hypernasality ratings for the most imprecise speakers are not reliable. While this is a limitation of the approach, it is not severely limiting. In most cases, clinicians are more concerned with evaluating speakers in the mild-moderate end of the scale where they can monitor disease progress early or evaluate the effects of an intervention. This is less common for later stages of disease.

It is interesting to note that despite poor alignment the model still yields high hypernasality scores for imprecise speakers. Precise alignment for speakers in this range is simply not possible, manually or otherwise. It's likely that the poor hypernasality ratings predicted by the model are driven by the poor alignment itself \cite{green2004revisiting}.

\subsection{\label{subsec:lowhpn} Analysis of Imprecise Articulation as a Confound}

In Section \ref{subsec:role} we analyzed the extent to which the NAP features genuinely capture both the perceptual dynamics of hypernasality and information beyond mere articulatory precision. In this section, we further analyze the role that articulatory precision alone may play as a confound, driving our system to label a low-hypernasality but otherwise severely dysarthric speaker as high hypernasality. After all, as discussed before, hypernasality and articulatory precision are well correlated---in our own data, the perceptual clinician AP and hypernasality ratings have a PCC of 0.84 to each other.

By dividing each speaker into differentiated classes based on the relationship between their own level of hypernasality and articulatory precision, we seek to assess if relatively low hypernasality but high CAP speakers are mislabeled as high hypernasality. To do this, we assess ``$\Delta_{Z}$,'' the difference between speaker $i$'s z-scored clinician AP and z-scored hypernasality scores ($CAP(i)$, $H(i)$),

\begin{equation}
    \Delta_{Z}(i)=\frac{CAP(i)-\mu_{CAP}}{\sigma_{CAP}} - \frac{H(i)-\mu_{H}}{\sigma_H},
\end{equation}
where the $\mu_{CAP}$, $\sigma_{CAP}$, $\mu_H$, $\sigma_H$ are the mean and standard deviation of the CAP and hypernasality scores, respectively.
 
We find that for our dataset, $\Delta_{Z}$ is roughly normally distributed, with a mean of $-0.001$ and a standard deviation of $0.57$, with a minimum of $-1.56$ and a maximum of $1.25$.

\begin{figure}[t]
\centering
    \includegraphics[width=\linewidth]{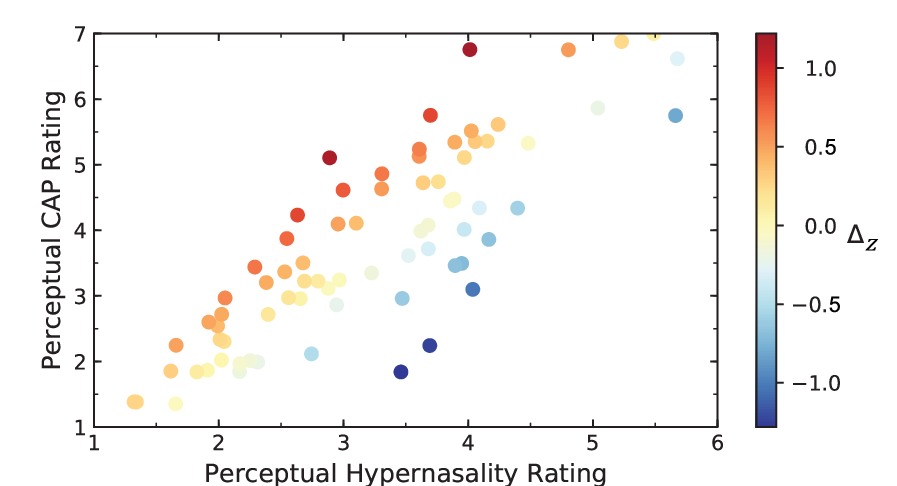}
\caption{A scatter plot of clinician-assessed perceptual hypernasality score against clinician articulatory precision rating (CAP), color-coded by $\Delta_{Z}$.}
\label{fig:cmap}
\end{figure}

Figure \ref{fig:cmap} shows a scatter plot of H against CAP for all speakers in our dataset, color-coded by $\Delta_{Z}$. F
or most levels of hypernasality or CAP, there exist a range of speakers with varying delta. Partitioning the $\Delta_{Z}$ distribution at $\pm0.3$ and $0$ splits it roughly in quarters.

To analyze the impact of imprecise articulation, we use $\Delta_{Z}$ and hypernasality scores to produce a train/test partition of the data, where the 7 speakers with $\Delta_{Z}>0.3$ and $H<2.5$, containing low-hypernasality speakers with relatively high CAP are the test set, while all others form the training set. 

We find that a linear model trained on the NAP features with this partition achieves an MAE of $0.823$ and a PCC of $0.560$. These results suggest that our model is robust to the correlation between hypernasality and the more generalized articulatory precision-reducing symptoms of dysarthria, as the model does not falsely label relatively high-CAP but low-hypernasality speakers as high hypernasality.

\section{Discussion}

The model based on NAP features outperforms both baselines, across all settings but MAE in the leave-out-HD case. Furthermore, the NAP-based model outperforms the human annotators in 8 of the 10 conditions, with the exception of LOSO PCC and leave-out-ALS PCC. In the LOSO condition the MFCC-NN approach outperforms the simpler formant features in PCC; while the formant feature model does achieve a lower MAE it seems to be a result of largely predicting the mean, with only a very modest upsloping trend in Figure 5(a) as opposed to Figures 5(b) and 5(c), which clearly show upward-sloping trends.

In the LODO conditions, the formant-feature-based and NN models perform unpredictably. On some disease classes, MFCC-NN outperforms FF, while the opposite is true for others. By comparison, the NAP achieves consistent performance across all LODO classes. This suggests that these features are a robust measure of hypernasality, relatively invariant to the disease-specific co-modulating variables that hinder the performance of the baselines on the same task. The nasalization features in the NAP are simultaneously more robust to both the disease-specific overfitting expected from NN methods such as \cite{hypernn} and speaker-to-speaker variances discussed in the design of the formant-based A1P0 and related features in \cite{styler}, \cite{chen1997}; this is the advantage of being trained on a large corpus of healthy speech, and targeting a specific perceptual quality. Articulatory precision features are robust in a similar way.

One of the added benefits of the proposed approach over the baseline methods is the direct interpretability of the individual NAP features. While it is not immediately clear how MFCC features or formant-based features are expected to change with different hypernasality levels, the proposed features are easy to interpret. 

The feature-level analyses of the nasalization and articulatory precision features behave as expected, with the nasalization log likelihood of the phonemes increasing as hypernasality increases, while the articulatory precision decreases as hypernasality increases (Figure \ref{fig:apf}).  Analysis of Eqns (3) and (4) shows that this makes sense. As hypernasality increases, the voiced phonemes become more and more like the $N$ class in the acoustic model in Section II. Similarly, as hypernasality increases, the acoustics of the unvoiced phonemes become less and less like the intended target, therefore the ratio in Eqn (4) decreases.

During the feature selection analysis in Section III-B, certain consonants appeared prominently. The nasalization feature for phonemes D, and B, as well as the articulatory precision of T and F were prominent. T, B, and D are referred to as a ``nasal cognates'' in \cite{ncd}, as the bilabial consonant B shares a place of articulation with the bilabial nasal M, the lingua-alveolar consonants T and D share a place of articulation with the lingua-alveolar nasal N. Leakage through the nasal cavity will interfere with the production of all of these phonemes, and in the voiced case, they will sound like their corresponding nasal phonemes. It is not surprising that the nasalization model is most sensitive to these phonemes since that model is trained on healthy speech, where the $N$ class consists mostly of instances in M and N and surrounding vowels.

Through the same analysis, the most prominent vowels selected were AA and IY. AA is the most open and back vowel in English, whereas IY is the most closed and fronted. It may be the case that these extreme ends of the vowel chart exhibit more noticeable patterns of nasalization, either on a perceptual level or just in their PLP-nasalization feature realization.

Dysarthria is characterized by a constellation of often co-modulating perceptual and acoustic attributes. Neuromotor speech control problems are not necessarily restricted to a specific articulator; when a speaker exhibits hypernasality arising from neuromuscular VPD, chances are other articulators such as the tongue and lips are affected, thereby driving a more generalized articulatory imprecision. The converse is also true, meaning that a speaker with some level of neuromuscular tongue, lip, and jaw-driven imprecision may also exhibit some amount of hypernasality. As a result it becomes difficult to truly isolate if a dysarthric hypernasality assessment system is precisely measuring hypernasality, rather than broader articulatory imprecision. To do this, experiments where only hypernasality varies as other articulatory imprecisions are held constant, and where articulatory precision varies against a low level of hypernasality are necessary. The former experiment is straightforward---carried out for our analysis on cleft palate speakers described in section \ref{subsec:role}. The latter was the experiment to evaluate the role that imprecise articulation plays as a confound in section \ref{subsec:lowhpn}.

We found that our hypernasality estimation model---trained exclusively on the dataset of dysarthric speech---achieved high correlation on estimating the hypernasality of CP speech. The CP recordings were all from speakers exhibiting no neuromuscular disease, meaning they exhibited hypernasality with no other articulatory imprecisions. Thus, the generalization results exhibited by our model suggest it models the genuine perceptual attributes of hypernasality in dysarthric speech, as opposed to the correlated generalized imprecision.

In spite of its robustness the NAP model has limitations. Most limiting is its reliance on aligned transcripts to perform the estimation. The results shown in this paper were based on forced alignment. This is always possible when the ground truth transcript is known but is not  feasible for spontaneous speech. The robustness of the model comes from the fact that it is trained on a large corpus of healthy speech; however, this training induces a bias in the model. As the feature selection results show, the model is adept at detecting hypernasal speech from phonemes that look similar to nasals in healthy speech; however it is impossible to capture nasalization acoustic patterns for unvoiced speech since these sounds never occur in healthy speech (and hence cannot be captured in our model). As a result, we use articulatory precision of unvoiced consonants as a proxy for nasalization for these sounds. Increased hypernasality typically implies reduced articulatory precision, but the converse is not necessarily true. As such, it is possible for speakers to exhibit reduced precision for other reasons than hypernasality. As we showed with the CLP speech experiments, when the reduction in articulatory precision is due to hypernasality, the model generalizes out-of-disease quite well. We did an initial analysis on 7 speakers with low hypernasality scores and relatively high clinical articulatory precision scores where the model showed that it tracks with hypernasality; however further analysis on larger dysarthric datasets are required to fully validate the model in this context.


 
 
 





\section{Conclusion}

We have presented and demonstrated the Nasalization-Articulation Precision features for objective estimation of hypernasality. This method leverages a data-driven approach to learning expert-designed features on healthy speech that capture perceptible elements in hypernasal speech. We demonstrated that these features, when evaluated on disordered speech, track the expected trends in perceptual hypernasality ratings, and can be used with ridge regression to estimate a clinician-rated hypernasality score more accurately than several representative baseline methods as well as human annotators, on average. Additionally, we demonstrated that the NAP algorithm predictions for hypernasality rating generalize across diseases with significantly less loss in accuracy than existing approaches. This implies that the NAP features are a robust method for estimating hypernasality in dysarthria.

The chief limitation of this approach, and articulatory precision estimation techniques more generally, is a reliance on known transcripts with which alignment may be performed. Neural models for directly assessing articulatory precision from raw speech audio is a promising future research direction---such models could provide the simultaneous identification of and precision assessment of phonemes on the fly, and provide downstream representations that could drive characterization of hypernasality without relying on reading as a stimulus, or known transcripts for assessment.

We plan to expand on this work by collecting a larger dataset of nasality-scored dysarthric speech, representing more diseases, and designing stimuli better tailored for this task. Furthermore, we will work to apply insights from this work to improve the robustness of neural models for the estimation of articulatory precision, nasality, and other objective speech biomarkers.

\ifCLASSOPTIONcaptionsoff
  \newpage
\fi



%
%

\bibliographystyle{IEEEtran}
\bibliography{bibliography}




\begin{IEEEbiography}[{\includegraphics[width=1in,height=1.25in,clip,keepaspectratio]{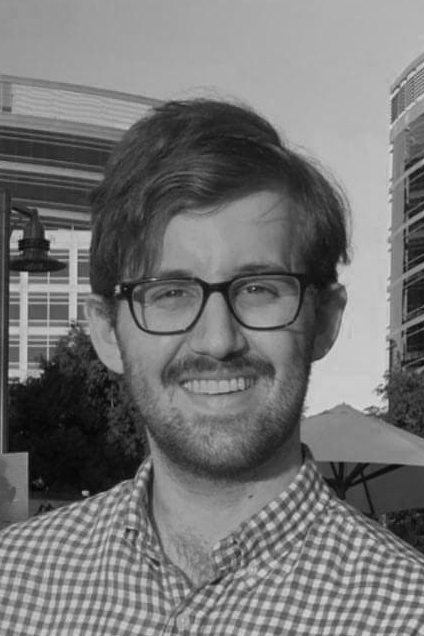}}]{Michael Saxon}
recieved the MS degree in Computer Engineering from Arizona State University (ASU) in Tempe, AZ, USA in May 2020, with a thesis topic centered on dysarthric speech analysis. He is now a Ph.D. student and NSF Fellow in the NLP Lab at the University of California, Santa Barbara (UCSB) in Santa Barbara, CA, USA. His research interests include automatic speech recognition, deep learning, and natural language processing.
\end{IEEEbiography}

\begin{IEEEbiography}[{\includegraphics[width=1in,height=1.25in,clip,keepaspectratio]{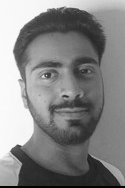}}]{Ayush Tripathi} 
received the B.Tech degree in Electrical and Electronics Engineering from Visvesvaraya National Institute of Technology, Nagpur, India in 2019. He is currently a researcher in the Speech \& Natural Language Processing Group at TCS Research \& Innovation, India. He has previously worked as a Research Aide at Arizona State University and a Research Intern at Indian Institute of Technology, Guwahati.
\end{IEEEbiography}

\begin{IEEEbiography}[{\includegraphics[width=1in,height=1.25in,clip,keepaspectratio]{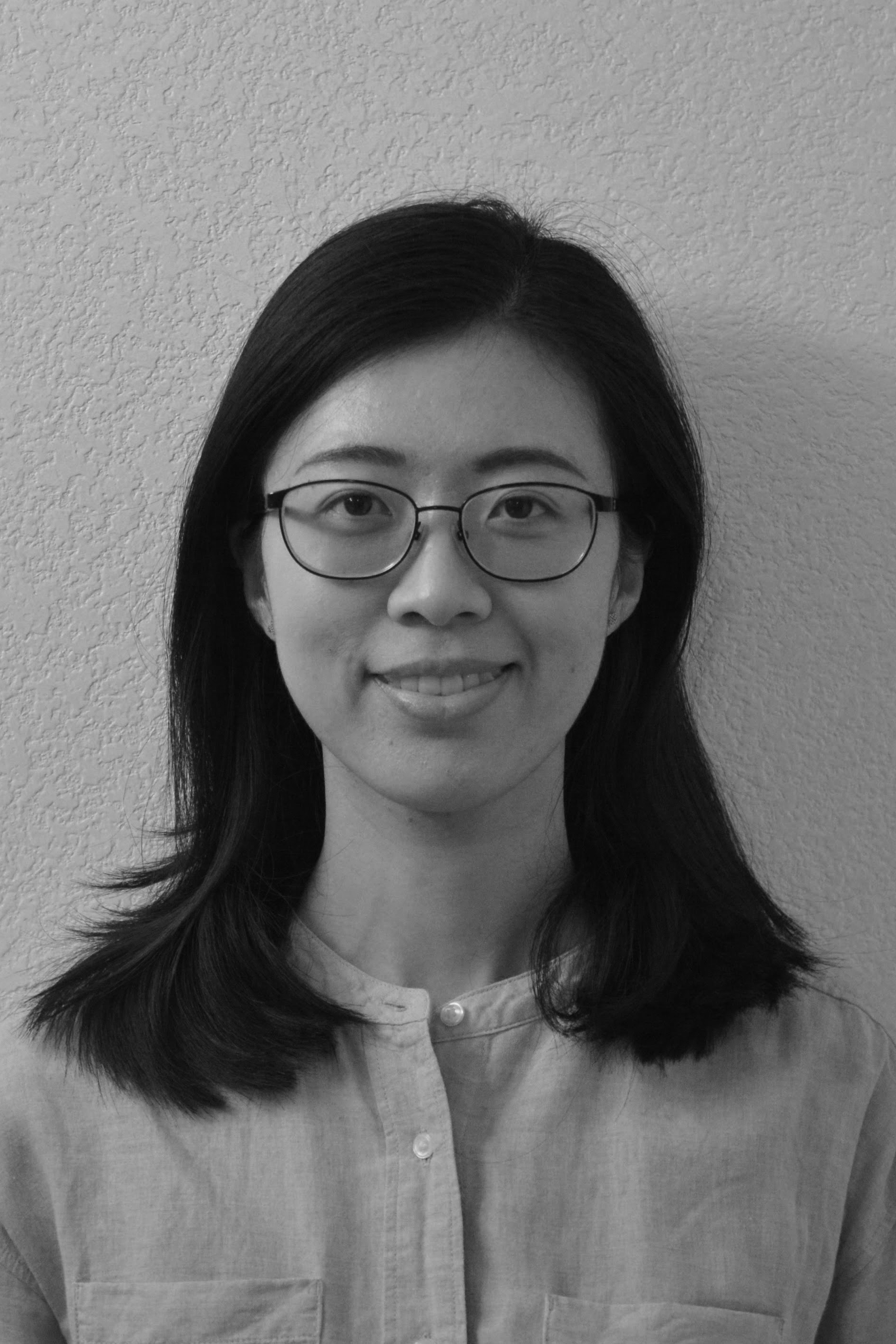}}]{Yishan Jiao}
received her Ph.D degree in Speech and Hearing Science from Arizona State University (ASU) in Tempe, AZ, USA in May 2019. Her research interests include acoustic signal processing and machine learning.
\end{IEEEbiography}

\begin{IEEEbiography}[{\includegraphics[width=1in,height=1.25in,clip,keepaspectratio]{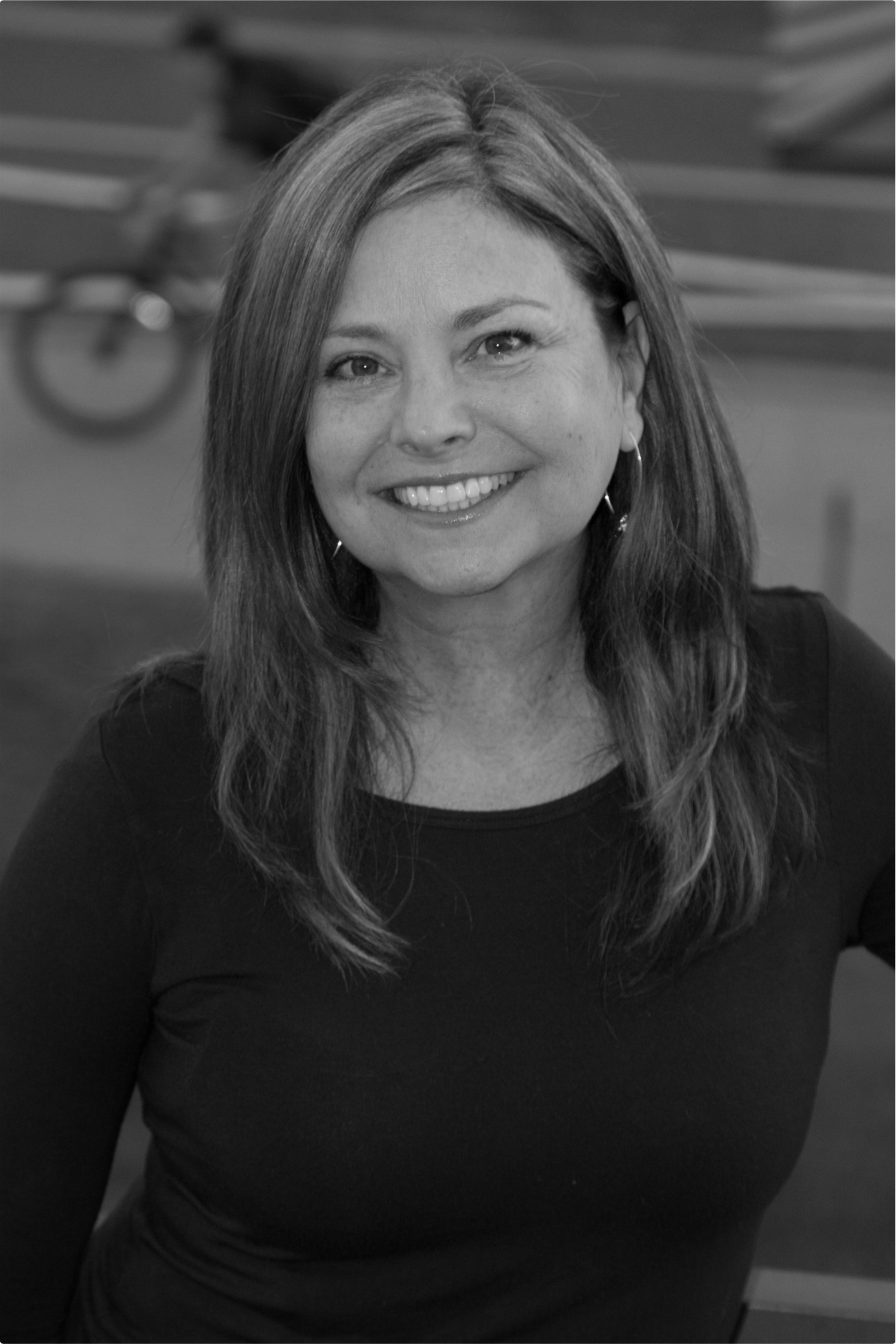}}]{Julie M. Liss}
is  a  Professor  of  Speech  \&  Hearing  Science  and Associate  Dean  of  the  College  of Health Solutions at Arizona State University (ASU) in Tempe, AZ, USA. Her research explores the ways in which speech and language change in the context of neurological damage or disease.\end{IEEEbiography}

\begin{IEEEbiography}[{\includegraphics[width=1in,height=1.25in,clip,keepaspectratio]{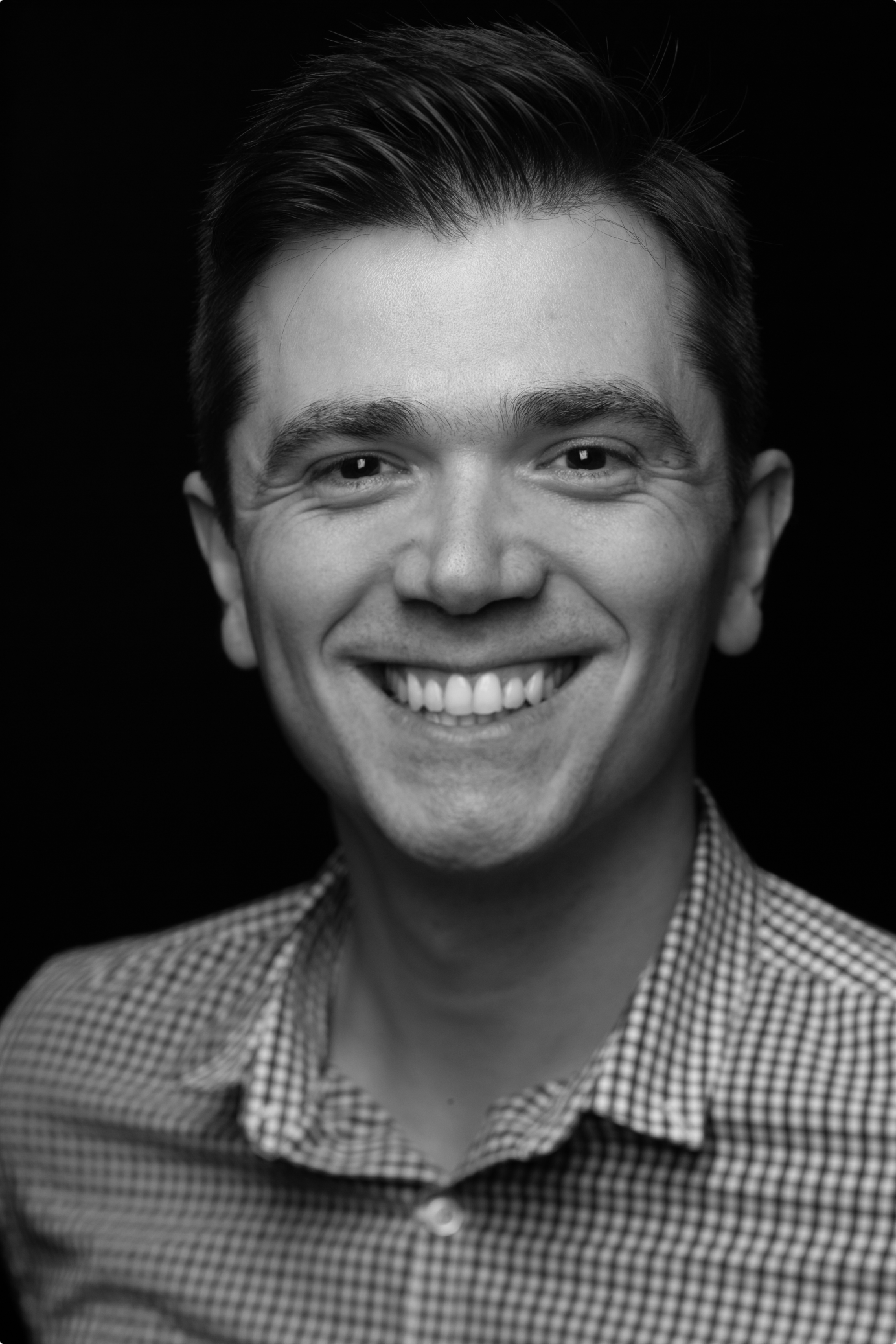}}]{Visar Berisha}
is an Associate Professor in the College of Health Solutions and Fulton Entrepreneurial Professor   in   the   School   of   Electrical   Computer \&  Energy  Engineering  at  Arizona  State  University (ASU)  in  Tempe,  AZ,  USA.  His  research  interests include computational models of speech production and perception, clinical speech analytics, and statistical signal processing.
\end{IEEEbiography}

\end{document}